\documentclass[preprint,5p]{elsarticle}

\usepackage{amssymb}
\usepackage{amsmath}
\usepackage{hyperref}
\usepackage{placeins} % in order to use floatbarrier

\newcounter{bla}

\journal{Computer Physics Communications}

\newcommand{\mpm}{\ensuremath{\pm}}
\newcommand{\vegasflow}{\texttt{VegasFlow}~}

\begin{document}

\begin{frontmatter}



\title{\texttt{VegasFlow}: accelerating Monte Carlo simulation across multiple hardware platforms}


\author[]{Stefano Carrazza\corref{author}}
\author[]{Juan M.~Cruz-Martinez}

\cortext[author] {Corresponding author.\\\textit{E-mail address:} stefano.carrazza@unimi.it\\\textit{Preprint number:} TIF-UNIMI-2020-8}
\address{TIF Lab, Dipartimento di Fisica,\\ Universit\`a degli Studi di Milano and
INFN Sezione di Milano,\\ Via Celoria 16, 20133, Milano, Italy}

\begin{abstract}
    We present \vegasflow, a new software for fast evaluation of high dimensional integrals based on Monte Carlo integration techniques designed for platforms with hardware accelerators.
    The growing complexity of calculations and simulations in many areas of science have been accompanied by advances in the computational tools which have helped their developments.
    \vegasflow enables developers to delegate all complicated aspects of hardware or platform implementation to the library so they can focus on the problem at hand.
    This software is inspired on the Vegas algorithm, ubiquitous in the particle physics community as the driver of cross section integration, and based on Google's powerful TensorFlow library. We benchmark the performance of this library on many different consumer and professional grade GPUs and CPUs.
\end{abstract}


\end{frontmatter}

\noindent
{\bf PROGRAM SUMMARY}
\\

\begin{small}
\noindent
{\em Program Title:} {\tt VegasFlow} \\
\\
{\em Program URL:} \url{https://github.com/N3PDF/vegasflow}\\
\\
{\em Licensing provisions:} GPLv3 \\
\\
{\em Programming language:} Python \\
\\
{\em Nature of problem:} The solution of high dimensional integrals require the implementation of Monte Carlo algorithms such as Vegas. Monte Carlo algorithms are known to require long computation times. \\
\\
{\em Solution method:} Implementation of the Vegas algorithm using the dataflow graph infrastructure provide by the TensorFlow framework. Extension of the algorithm to take advantage of multi-threading CPU and multi-GPU setups.\\

\end{small}

%\tableofcontents

\section{Introduction and motivation}
\label{sec:introduction}

State-of-the-art computations in High Energy Physics (HEP) require computing very complex multi-dimensional
integrals numerically, as the analytical result is often not known.
Monte Carlo (MC) algorithms are generally the option of choice, be it in the HEP application or elsewhere,
as the error of such algorithms does not grow with the number of dimensions.

In particular, in the HEP literature, MC methods based on the idea of importance sampling are widespread
as they combine the robustness of MC algorithms for high dimensional situations with the flexibility of
adaptative grids.

The Vegas algorithm~\cite{Lepage:1977sw, Lepage:1980dq}
is the main driver for multi-purpose parton level event generation programs based on fixed order calculations
such as MCFM\cite{Campbell:2015qma, Campbell:2019dru}, NNLOJET~\cite{Gehrmann:2018szu}
and also of more general tools such as MG5\_aMC@NLO~\cite{Alwall:2014hca} and Sherpa~\cite{Gleisberg:2008ta}.
Whereas the original implementation of the algorithm was written for a single CPU, nowadays it is usually
implemented to take advantage of multi-threading CPUs and distributed computing.
Indeed, MC computation are what is informally known as ``embarrassingly parallel''.

However, the parallelization of a computation over multiple CPUs does not decrease the number of CPU-hours
required to complete a computation and the cost of such calculations is driving the budget of
big science experiments such as ATLAS or CMS~\cite{Buckley:2019wov}.

In this paper we present the \vegasflow library~\cite{juan_cruz_martinez_2020_3691927}, where the main contribution is a novel implementation
of the importance sampling algorithm used in the aforementioned event generation programs able to run
both in CPUs and GPUs, enabling further acceleration of complicated integrals.
The library is written using the TensorFlow~\cite{tensorflow2015:whitepaper} library hence the chosen name: \vegasflow.

With this publication we do not aim to overthrow or dethrone Vegas but rather
empowering it even more by enabling the frictionless integration of complicated processes
in all kinds of hardware supported by TensorFlow with little to no effort made by the user.
% should we add some comment here about how important this part is?
% but I don't want to offend anyone...

The importance sampling ``\`a la'' Vegas is the main algorithm included in \vegasflow but the library is
designed such that new algorithms can be easily implemented.
We believe this design choice together with the TensorFlow back-end will enable a much faster development
cycle towards the much desired goal of a Neural Network-based integration algorithm able to surpass Vegas
and further reduce computational costs.
This feat is yet to come and the effort is not limited to the HEP community but it is rather
multidisciplinary.
One such example is the Neural Importance Sampling~\cite{DBLP:disneypaper} developed in the context
of image rendering whose finding have inspired new research in the field of particle physics~\cite{Gao:2020vdv,Gao:2020zvv,Bothmann:2020ywa}.

\section{Technical Implementation}

The goal of this manuscript is to present a novel open-source library for Monte Carlo integration which takes advantage from hardware accelerators such as GPUs, lowering the barrier in terms of computational knowledge from the user point of view.
Our motivation is primarily technical as until now there are no public available libraries which provide such features and thus we think that the scientific community may benefit from a practical implementation. Our aim is for \vegasflow to set a new implementation standard for future Monte Carlo calculations.

\subsection{Acceleration paradigm}

Hardware acceleration combines the flexibility of general-purpose processors, such as CPUs, with the efficiency of fully customized hardware, such as GPUs, ASICs and FPGAs, increasing efficiency by orders of magnitude. In particular, hardware accelerators such as GPUs with large number of cores and memory are getting popular thanks to its great efficiency in deep learning applications through open-source frameworks such as TensorFlow which simplifies the development strategy by reducing the required hardware knowledge from the developer point of view. In this context, \vegasflow implements for the first time a Monte Carlo integration produce using TensorFlow primitives together with job scheduling for multi-GPU synchronization. The choice of TensorFlow as the back-end development framework for \vegasflow is motivated by its simple mechanism to write efficient python code which can be distributed to hardware accelerators without complicated installation procedures.

\subsection{Integration algorithms}

The main algorithm in VegasFlow is importance sampling as implemented in Vegas~\cite{Lepage:1977sw, Lepage:1980dq}, hence the name
chosen for the library.

Nonetheless, the library aims to be a general purpose MonteCarlo library.
We provide a \texttt{MonteCarloFlow} class from which the developer can inherit in order to construct a custom integrator
algorithm.
The developer has to worry just about what the integrator should do for every particular event (for instance, how to
generate the random numbers) and what to do after an integration is finished (for instance, refine how the random
numbers are generated).
All other technicalities, GPU distribution, multithreading or vectorization of the computation will be dealt with
by the library.

\subsection{Integrands}

For a better integration with VegasFlow, integrands should be written with TensorFlow primitives in python.
Written operations using TensorFlow operators allows for the usage of all the hardware TensorFlow is compatible with.
The library, however, is not limited and can run integrands written in Fortran, C/C++ or even CUDA~\cite{nickolls2008scalable} through the CFFI library~\footnote{\url{https://github.com/cffi/cffi}}.
Alternatively, both C++ and CUDA integrands can be easily linked as TensorFlow operators. The \vegasflow package available in~\cite{juan_cruz_martinez_2020_3691927} contains some examples in the source code.

Features such as exporting histograms during integration can also be implemented and some examples are packaged with the source code.

\begin{table*}
    \centering
    \begin{tabular}{ c | c  c  c c }
       Integrand                           & Plain MC       & Vegas                          & VegasFlow CPU                & VegasFlow GPU\\ \hline
SymGauss 8-d                        & 0.99 \mpm 0.08 & 1.00002 \mpm 0.00023 (18.7s)   & 1.00005 \mpm 0.00018 (9.87s) & 1.00008 \mpm 0.00016 (6.21s) \\
SymGauss 20-d                       & -                & 1.00003 \mpm 0.00002 (38min) & 1.00002 \mpm 0.00005 (26min) & 1.00003 \mpm 0.00003 (5min) \\
Genz Eq.\eqref{eq:genzdisc} 16-d & -                   & 0.99992 \mpm 0.00008 (1004s)   & 1.00010\mpm 0.00011 (609s)   & 0.99998 \mpm 0.00009 (86s) \\
Genz Eq.\eqref{eq:productpeak} 16-d & -               & 0.99996 \mpm 0.00011 (1086s)   & 1.00013 \mpm 0.00010 (468s)  & 1.00026 \mpm 0.00020 (92s) \\
       \hline
    \end{tabular}
    \caption{Comparison of \vegasflow with other MC implementations. The number of events per iteration is
    constant for all integrators for a given integrand. The Plain MC is able to get results in a reasonable amount of time only for the Symmetric Gaussian function in 8 dimensions. The same feature is observed for all integrands where the GPU run of \vegasflow achieves the final result much faster than its CPU or Vegas~\cite{peter_lepage_2020_3647546} counterparts. Note that the choice of parameters (\textit{e.g.} number of subdivisions of the grid) are arbitrarily chosen for the purposes of this benchmark to be the same across implementations but are not necessarily the optimal choices.}
    \label{table:precision}
\end{table*}

\section{Benchmark}

\subsection{Toy integrands}
As a first test and benchmark of \vegasflow we use several toy models for which the analytical solution is known.
We start by using a spherically symmetric Gaussian as it was also the first example shown in the original Vegas paper~\cite{Lepage:1977sw}.
\begin{equation}
    \label{eq:lepage}
    I_{n} = \mathcal{N} \exp \left[-\frac{1}{a^2}\displaystyle\sum_{i=1}^{n}\left(x_{i}-\frac{1}{2}\right)^{2}\right].
\end{equation}

We also implement some of the integrands proposed by Genz as a test of multidimensional integration algorithms~\cite{genzint}, in particular the Genz discontinuous function
\begin{gather}
    I_{n} = \left\{
\begin{array}{ll}
    0 & \text{If any } x_{i} \leq a_{i} \\
    \mathcal{N} \exp \left(\displaystyle\sum^{n}_{i=1} x_{i}c_{i}\right) & \text{otherwise} \\
\end{array}
\right. \label{eq:genzdisc}
\end{gather}
and the Genz product peak function
\begin{equation}
    I_{n} = \mathcal{N} \displaystyle\prod_{i=1}^{n} \frac{1}{c_{i}^{-2} + (x_{i}-a_{i})^{2}}. \label{eq:productpeak}
\end{equation}

In  all cases the factor $\mathcal{N}$ normalizes the integrand such that it integrates exactly to one. The number of dimensions is set by $n$ and the difficulty of the integration increases with $c_{i}$.

\subsection{Eager mode vs graph mode}

In TensorFlow 2 the so called eager execution was introduced as the default behaviour.
Eager execution implements the imperative programming paradigm into
TensorFlow, and as a consequence, statements are executed in place instead
of building a graph that is subsequently executed later in the program.
In this mode, the development and debugging is simplified in exchange for an expected decreased performance.

\begin{figure}
    \center
    \includegraphics[width=0.48\textwidth]{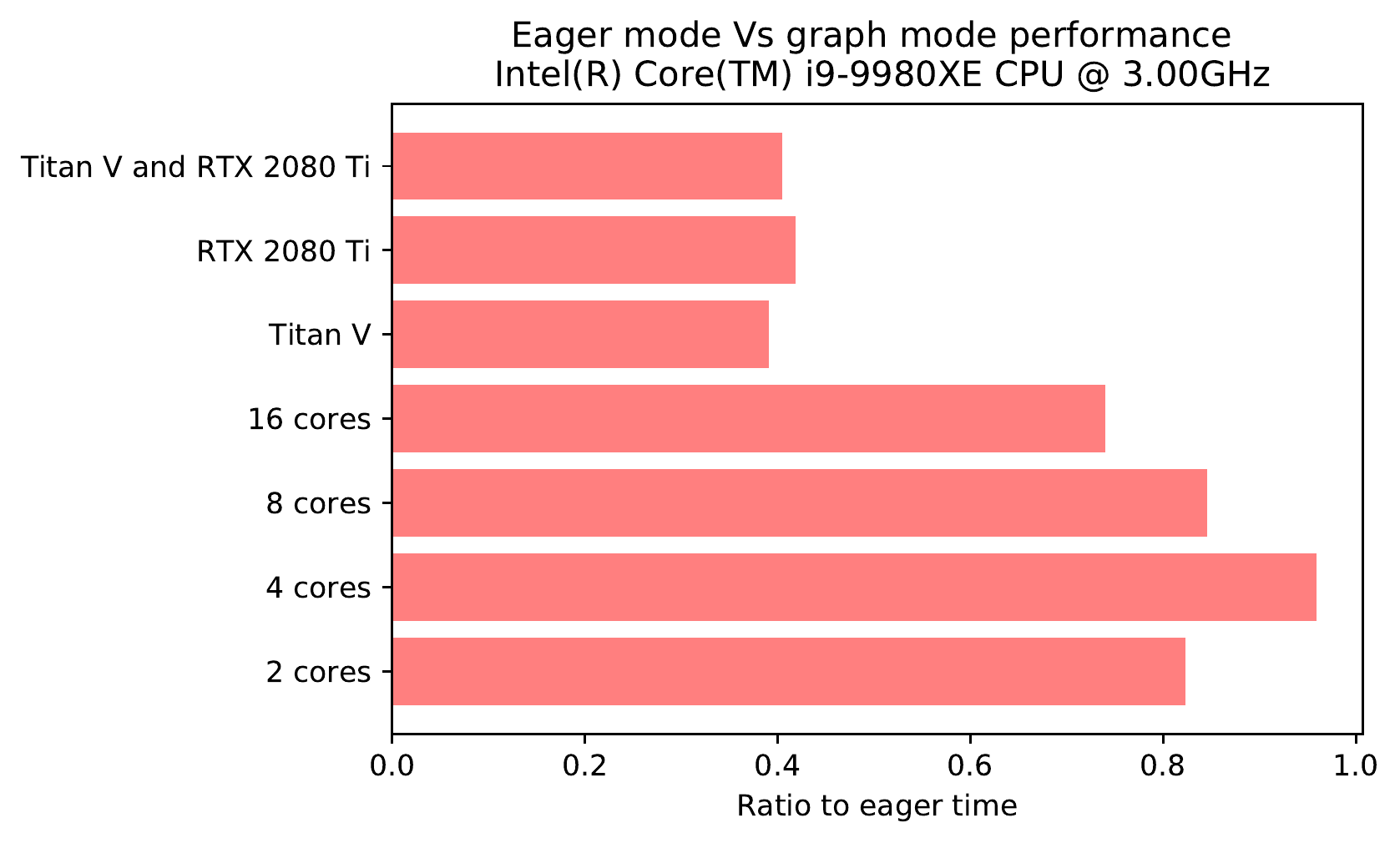}
    \caption{Comparison of performance between the eager and graph compilation TensorFlow mode. The results
    are shown as a ratio of the time it took the eager computation to complete one iteration. We find comparable
(albeit improved) results when running the compiled graph in only CPU mode but a 3x improvement when running on the GPUs.}
    \label{fig:eagerperformance}
\end{figure}

In order to quantify the performance hit of the eager mode in comparison to the graph mode we run the same integration in both modes and show the results in Fig.~\ref{fig:eagerperformance}.
In such figure we compare the results of a professional-grade CPU (Intel i9-9980XE) with a consumer-grade GPU (NVIDIA RTX 2080 Ti) and a professional-grade GPU (NVIDIA Titan V).
We find the greater improvement with respect to eager execution is found in highly parallel scenarios, such as multi-CPU computation or GPU runs.

It is clear that, whereas eager mode facilitates development, production runs of the code should always be run on graph compiled mode.

\subsection{Result benchmark}

As a first test we ensure that our integrator produces the correct results for several different integrands.
For this we make use of Lepage's python implementation of the Vegas
algorithm~\cite{peter_lepage_2020_3647546} and a plain MC algorithm with no adaptation.

The results are shown in Table~\ref{table:precision}.
It can be observed that both implementations of importance sampling produce (as one would expect) compatible result.
Both Vegas and \vegasflow CPU are using all CPUs from an Intel(R) Core(TM) i9-9980XE CPU.
In the next section we perform a more detailed benchmark of the running time of different integrators but we can already
see a strong improvement due to the usage of the GPU by \vegasflow as the computation becomes more complicated.

\subsection{Performance}

Arguably the main contribution from \vegasflow is the ability to use one single implementation across many different devices.
The GPUs can take advantage of the huge parallelizability of Monte Carlo simulations reducing the time it takes to finish one computation in an order of magnitude.
Furthermore, the reduction is also very apparent on the power consumption of the different devices, indeed, as seen in
Table~\ref{table:cpuvsgpupower}, running the same computation is much more slow and expensive when it is run in the CPU.
The average power consumption of the CPU is comparable to the Titan V, but with a much longer computational time.

\begin{figure}
    \center
    \includegraphics[width=0.435\textwidth]{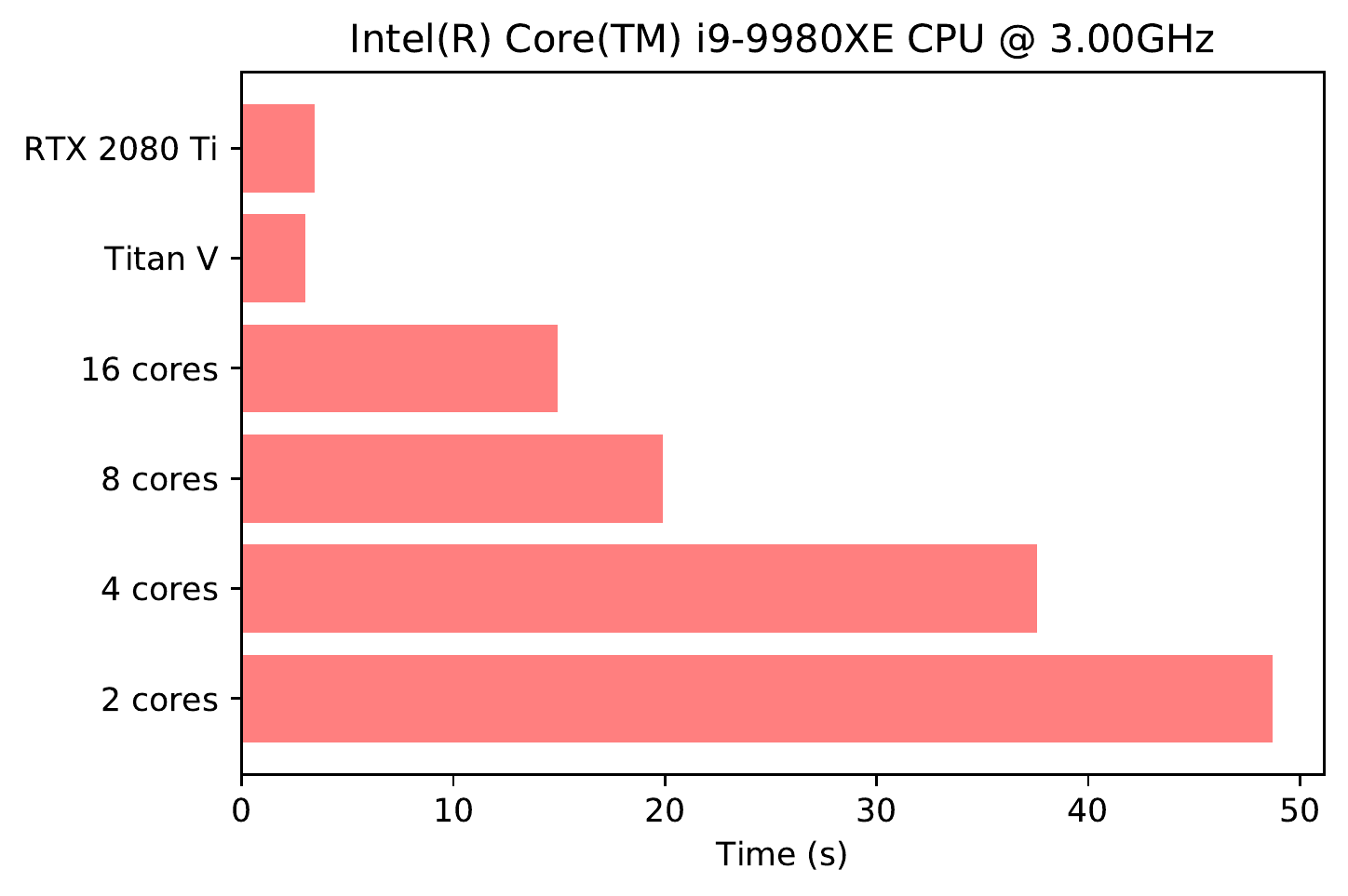}
    \caption{\vegasflow running the same integration in 2, 4, 8, 16 threads in a CPU and in the RTX 2080 Ti and Titan V
    GPUs. Less is better.}
    \label{fig:cpuvsgpuperformance}
\end{figure}

\begin{table}[ht]
    \centering
    \begin{tabular}{c c c} \hline
        Device        & Total Time   & Avg. Power Consumption \\ \hline
        i9 (16 cores) & 609s & 85 W  \\
        RTX 2080 Ti   &  93s & 105 W \\
        Titan V       &  89s &  75 W \\ \hline
    \end{tabular}
    \caption{Comparison on the power consumption of different devices. The CPU power consumption is provided by the
    \texttt{powertop} utility while the GPU power consumption is a sum of the power draw reported by \texttt{nvidia-smi} and
\texttt{powertop}.}
    \label{table:cpuvsgpupower}
\end{table}

\begin{figure}[t]
    \center
    \includegraphics[width=0.49\textwidth]{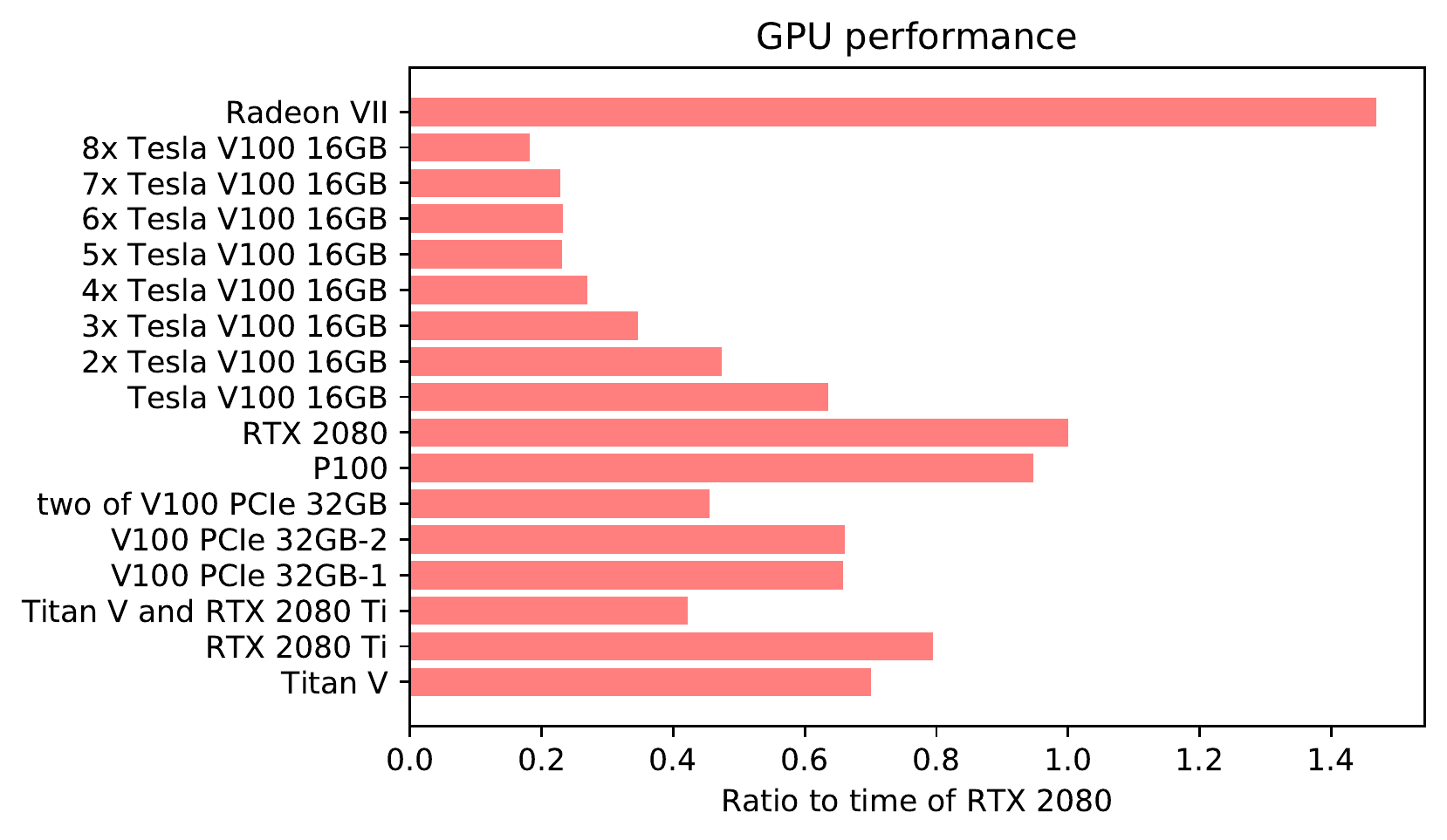}
    \caption{\vegasflow running in different GPU devices. We use the consumer-grade RTX 2080 as the measure of
    performance (less is better).}
    \label{fig:gpuperformance}
\end{figure}

\begin{figure*}
    \center
    \includegraphics[width=1.0\textwidth]{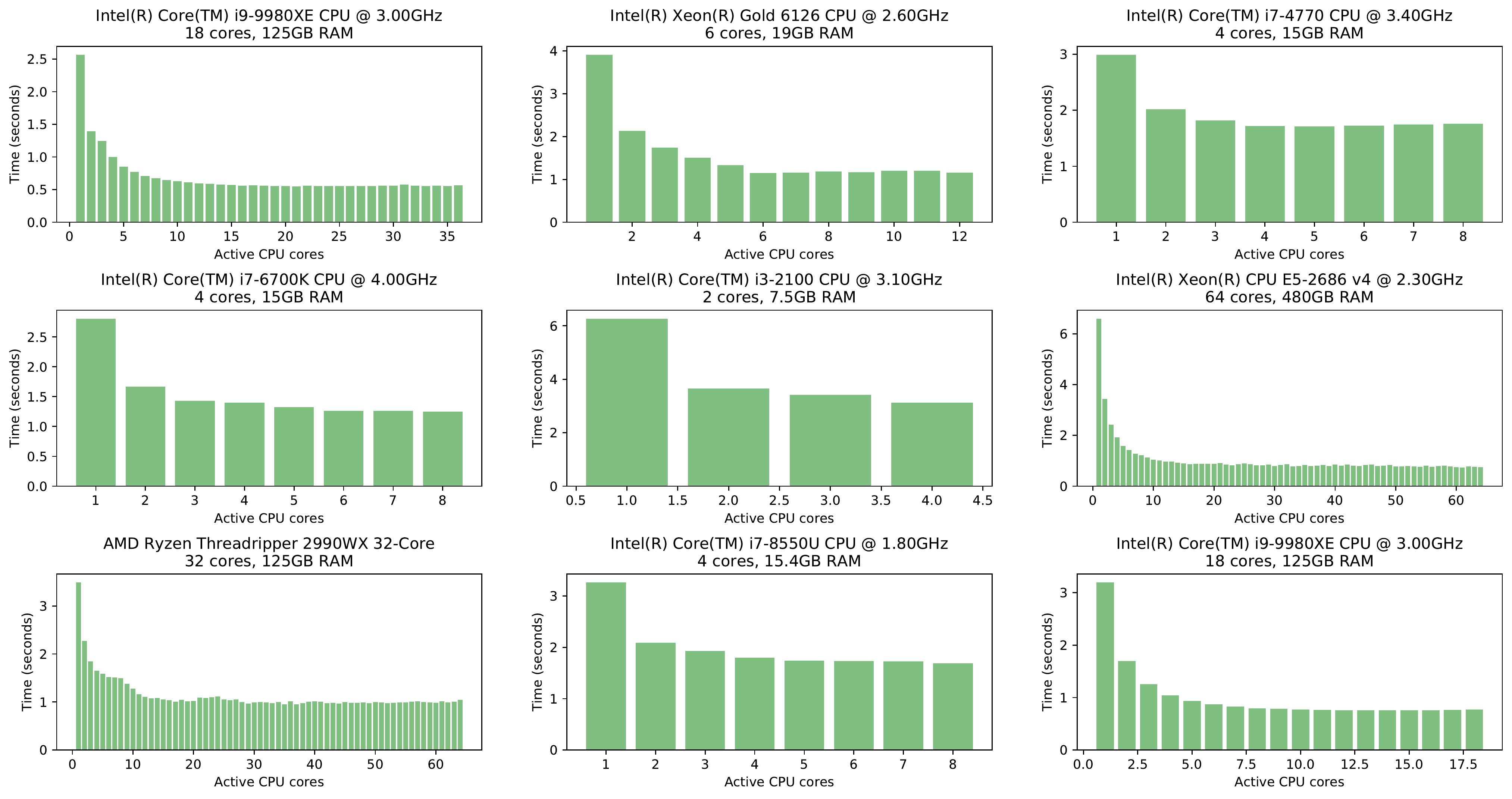}
\caption{Benchmark of \vegasflow on CPU. We observe an improvement of the performance as the number of allowed cores grows until the number of allowed cores is of the same order of the number of physical cores in the machine. The control of the threads for each run was left to TensorFlow with default values, while the binding of the process to the desired number of cores was done with \texttt{taskset}.}
    \label{fig:cpuperformance}
\end{figure*}

We have also made sure \vegasflow can be used in a multi-GPU setting with many different brand and devices.
At the moment only distribution on devices within the same physical machine is supported but we plan to implement distribution over different physical machines.
We can observe in Fig.~\ref{fig:gpuperformance} how the hierarchy between different GPU devices corresponds to what one would expect from their technical specifications.
In Fig.~\ref{fig:gpuperformance} we also observe a good scaling of the speed of the code with the number of GPUs although as the number of GPUs grow we hit a problem of diminishing returns. Similar results are also presented for different CPU models in Fig.~\ref{fig:cpuperformance}.

\begin{figure}
    \center
    \includegraphics[width=0.48\textwidth]{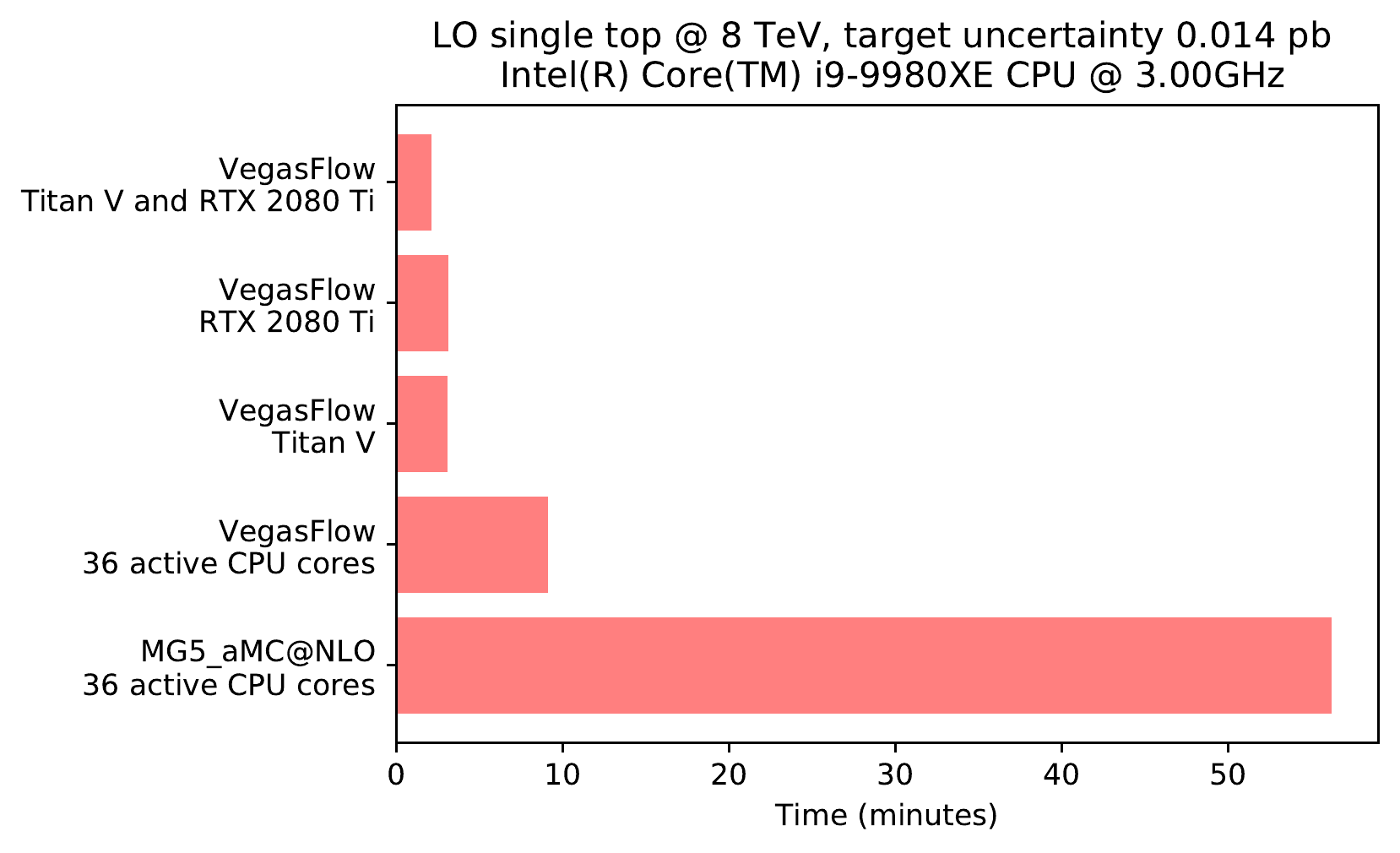}
    \caption{Comparison of a Leading Order calculation ran in both \vegasflow and MG5\_aMC@NLO~\cite{Alwall:2014hca}. The CPU-only version of \vegasflow is able to improve the performance obtained by MG5\_aMC@NLO for the same level of target accuracy. The usage of the GPU devices further improves the performance.}
    \label{fig:mg5}
\end{figure}

\subsection{Single $t$-quark production at leading order}

For the purposes of this benchmark we have considered the calculation of a particle physics process at the partonic level,
this is, without considering the convolution with the parton density functions (PDFs).
We compare our calculation with the numbers produced by MG5\_aMC@NLO~\cite{Alwall:2014hca} for the single $t$-quark production (t-channel) at leading order (LO)~\cite{Brucherseifer:2014ama} using the same physical parameters such as the $t$-quark mass, $m_t=173.2$ GeV and centre of mass energy $\sqrt{s}=8$ TeV.

In Fig.~\ref{fig:mg5} we compare the execution time for \vegasflow for the single GPU, multi-GPU and multithreading CPU configurations with the equivalent fixed LO order provided by MG5\_aMC@NLO 3.0.2. The stopping criteria for the total number of events relies on a target accuracy of $1.4\cdot10^{-2}$ pb (with no PDFs). Finally, also in this setup, we observe a great improvement in terms of execution time for the \vegasflow approach.

\section{Outlook}

We hope this library can accelerate research by granting to users and researchers the ability to implement with simplicity high-dimensional complex integrations without having to know about the technicalities or the difficulties of their implementation on multithreading systems or the data placement and memory management that GPU and multi-GPUs computing requires.

\vegasflow is also aimed to developers of new integration methods which can focus on the algorithm technicalities
and reduce to a minimum the implementation effort required to adapt the computation into different hardware platforms.
% as Steve Ballmer would say: developers, developers, developers, developers

The current release of \vegasflow has only been tested in GPUs and CPUs, however we believe that investigation about new hardware accelerators such as Field Programmable Gate Arrays (FPGA) and Tensor Processing Units (TPUs) could provide even more impressive results in terms of performance and power consumption results.

\section*{Acknowledgements}

We thank Stefano Forte for a careful reading of the manuscript.
We thank Durham University's IPPP for the access to the P100 and V100 32 GB GPUs used in order to benchmark this code.
We also acknowledge the NVIDIA Corporation for the donation of a Titan V GPU used for this research.
This project is supported by the European Research
Council under the European Unions Horizon 2020 research and innovation Programme (grant agreement number 740006) and
by the UNIMI Linea2A project ``New hardware for HEP''.


\section*{Bibliography}
\bibliographystyle{elsarticle-num}
\bibliography{../blbl}






% \end{thebibliography}

\end{document}